# Resonant structure for improved directionality and extraction of single photons

SAGAR CHOWDHURY,[1] RITURAJ,[2,] SRINI KRISHNAMURTHY[1,3], AND VIDYA PRAVEEN BHALLAMUDI[1*]

[1]*Dept. of Physics and Quantum Center of Excellence for Diamond and Emerging Materials (QuCenDiEM), Indian Institute of Technology Madras, India*
[2]*Indian Institute of Technology Kanpur, India*
[3]*Sivananthan Laboratories, Bolingbrook, IL, USA*
\*: Corresponding author, praveen.bhallamudi@iitm.ac.in

**Abstract:** Fluorescent atomic defects, especially in dielectric materials, such as diamond are quite promising for several emerging quantum applications. However, efficient light extraction, directional emission, and narrow spectral emission are key challenges. We have designed dielectric metasurface exploiting Mie-resonance and the Kerker condition to address these issues. Our designed diamond metasurface, tailored for nitrogen-vacancy (NV) defect centers in diamond, predicts up to 500x improvement in the collection of 637 nm (zero phonon line) photons over that from the bare diamond. Our design achieves highly directional emission, predominantly emitting in a 20° lobe in the forward direction. This makes light collection more efficient, including for fiber-based collection. The predicted results are stable against the position of the emitter placed in the metaelement, thus alleviating the challenging fabrication requirement of precise positioning of the defect center. Equally importantly, our design approach can be applied to enhance single photon emission also from other defects such as SiV, other materials such as hBN, and other sources such as quantum dots.

### 1. Introduction

Defect centers in wide bandgap materials, such as diamond, hexagonal boron nitride (hBN) and silicon carbide (SiC), have shown tremendous promise for both photon and spin-based quantum technologies, which can be operated at room-temperature. Several single photon emitting defects have been identified in these materials [1–4]. Single photon sources are essential for several quantum technologies, but high brightness and truly single photon emitters are still an open challenge [5]. Similarly, defect-based single spin centers have been established in these materials, and they typically rely on optical excitation and read-out using fluorescence.

Many of the demonstrations have used the well-explored color center—the negatively charged nitrogen-vacancy (NV) center in diamond. It consists of a substitutional nitrogen atom with a nearest neighbour carbon vacancy. It exhibits long electronic spin coherence time of ~ms even at room temperature and has been used to demonstrate quantum registers/memories [6] and highly sensitive nanoscale magnetometry [7].

However, photon collection from NV centers and other defects in diamond are limited by several factors. First, owing to the large refractive index (2.42) of diamond, the Fresnel-limited photon extraction efficiency is low. From emitters situated inside a bulk diamond (even at 100 nm below from the top surface), only <5% of the total emitted light comes out of a planner surface. Second challenge is that the emission from the atomic defect is isotropic, which makes the collection efficiency low even with high numerical aperture (NA) lenses. Third, the emission spectrum of NV center is broad because of broad phonon-side band [Figure S1], resulting in only 3-4% of the emitted light at the zero-phonon line (ZPL) wavelength of 637 nm [8]. This makes the system less suitable for single photon applications and for spin-photon hybrid quantum applications [9–11].

Structural modification of the bulk sample can address several of these issues. Different types of photonic structures have been fabricated with the aim of addressing the above-mentioned challenges. Fabrication of hemispherical solid immersion lenses centered around the color center allowed the photons to exit at normal angle and thus increased the extraction by almost ten times [12]. However, this emission into air is uniformly over $2\pi$ solid angle. The nanopillars or nanowires act as waveguide and allow ~ 40% of the emitted light to come out, but the emission is over both forward and reverse directions (into the bulk) [13]. Even more importantly, these non-resonant structures do not address the issue of broadband emission. The Purcell effect has been successfully employed to enhance the spontaneous emission at ZPL [14]. By coupling single NV center to a 2D photonic crystal structure made of diamond, the emission at ZPL has been enhanced by 70-fold [15]. However, these designs required an additional external coupler such as grating coupler to achieve partial directionality [16] [17].

In this article, we describe a structure and a design methodology to achieve both enhanced emission and directionality at a desired wavelength. We designed Mie-resonant [18] [19] nanopillars array, around an emitter, that meets Kerker condition [20–22]. Our aim is to increase the ZPL photons from NV in diamond. We show a design that is predicted to achieve ZPL photon counts which is >200x enhancement over that from bulk diamond. Further the emission is restricted to within a 20° cone in forward direction. The design principle and performance can be extended to other emitters in diamond as well as to other materials such as hBN, SiC, and other sources such as quantum dots.

The rest of the paper is organized as follows. In Sec. 2, we discuss our proposed structure, the design principle, and the algorithm to find the optimized structural parameters for a chosen material and wavelength. We provide detailed discussion on the properties of the supported electromagnetic modes that enable the Kerker condition using modal decomposition method. In Sec. 3, we discuss our predicted results of enhanced emission, directionality, and spectral selectivity. The concluding remarks are given in Sec. 4.

## 2. Design principles and algorithm

Our design of diamond metasurface containing NV center uses shape-dependent Mie resonance to enhance the electric field strength inside the meta-element—which effectively increases the photon emission by Purcell effect— and Kerker condition to enhance forward emission. When the wavelength of light in the metaelement is comparable to the size of the metaelement, a standing wave is formed and the electric field strength inside the element increases considerably. When a dipole emitter is placed in the element, the light matter interaction is increased, and the brightness (number of emitted photons) increases. Further, as the photonic density of states outside the resonance wavelength is typically reduced the photon emission at these undesired wavelengths is suppressed, resulting in further increase in emitted ZPL photon count. However, unless designed carefully, the emission will be in both forward and reverse direction. The Kerker condition—appropriate choice of modes to destructively interfere in the backward emission direction— is used to suppress the emission in the reverse direction, resulting in increased forward emission. Optimizing the size of the structures allows the interference between the radiation patterns of the different modes to fully suppress either forward or backward emission. In our design we show that the condition can be achieved through the interference between the electric dipole (ED) and magnetic dipole (MD) modes under the excitation of an electric point dipole (defect center) inside the pillar. However, note that a strong Mie resonance with large Purcell factor requires the medium to have larger refractive index $n$ for strong electric field confinement, and we need to bring different modes at the resonance to satisfy the Kerker condition. To satisfy both conditions, a proper design and optimization principle is required.

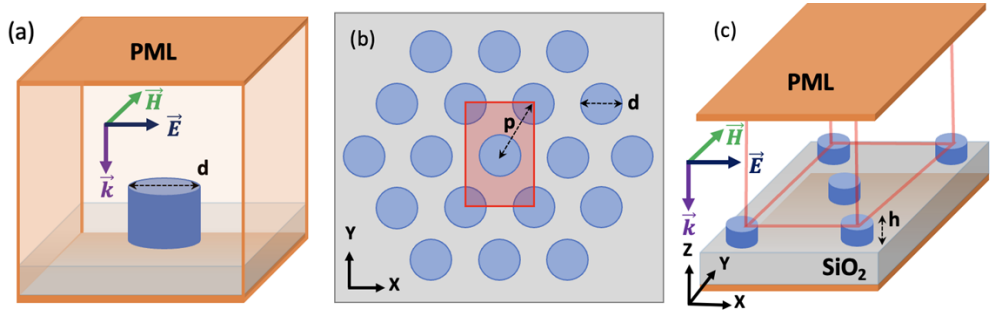

Fig. 1. **Design algorith and unit cell of diamond based metastuctures**. (a) Unit diamond nanopillar is placed on the glass substrtae, is excited under plane wave source to find the optimized diameter d. (b) Two dimentional top view of the pillars in the hexagonal lattice arry of periodicity p. (c) Unit cell of the hexagonal lattice, where pillars are placed on glass, perfectly matched layer (PML) is used at the bototm and the top of the unit cell. Red lines are correponds to periodic boundary condition (PBC) to consider an infinite array.

We chose metasurface made of diamond nanopillar of diameter $d$ and height $h$ — arranged in a hexagonal lattice array of periodicity $p$—on a low-index transparent (at ZPL) dielectric substrate for higher index contrast and thus better mode confinement inside the nanopillars as shown in Figure 1. The values of $n$ for substrate (silica/quartz) and diamond are 1.458 and 2.42 at 637 nm [23]. The design optimization is accomplished with Lumerical finite-difference time-domain (FDTD) solver [24], by varying $h$, $d$, and $p$ to maximize the local photon density of states around 637 nm. The optimization algorithm proceeded as follows:

First, we optimize $d$ by considering a single diamond nanopillar with unit aspect ratio ($h=d$) as shown in Figure 1(a). The perfectly-matched-layer or PML boundary condition is used at all the boundaries of the unit cell to simulate semi-infinite layers. We calculated the three-dimensional electric field profile in the pillar for different values of $d$ and $h$ by launching a plane polarized broadband light source from the top of the pillar. Through the multipole expansion method [25], we decomposed the contribution of ED, MD, and electric quadrupole (EQ) and magnetic quadrupole (MQ) modes. We observed maximum ED contribution to the scattering cross-section at 637 nm corresponding to diameter of 320 nm. The ED and other modes extend over a broad wavelength range, due to moderately low index value of diamond, which results in poor electric field confinement and thus low-quality factor [Figure S2]. In our design, better confinement is provided by the lower index substrate.

Second, after optimization of a single nanopillar, we considered the collective Mie resonances of a periodic array of diamond pillars. The array enhances the Mie resonance while allowing the modes to mix and satisfy the Kerker condition. We optimized the lattice period $p$ to achieve a better mode confinement and more directional radiation pattern. We observed multiple resonance peaks in reflection and transmission spectra as a result of interacting Mie resonances. The linewidth of the resonance peaks decreases when the periodicity is reduced, indicating resonance peaks in reflection and transmission spectra as a result of interacting Mie resonances. The linewidth of the resonance peaks decreases when the periodicity is reduced, indicating stronger field confinement and higher Purcell factor inside the nanopillars [Figure S3]. However, for fabrication considerations, we chose a value 410 nm for $p$ for the optimized value of 320 nm for $d$.

Third, we then optimized $h$ of nanopillars for the above chosen values of $p$ and $d$. As before, the electric field profile and the associated modes, deduced from multipole expansion method, are identified for each value of $h$. We found the electric field magnitude to be maximum for nanopillar height of ~550 nm [Figure S4] which results from a dominant ED resonance. We observed the ED resonance at the wavelength of 600 nm for the chosen values of the parameters

[Figure S5], and then by simple scaling of all parameters the ED resonance is moved to 637 nm [26]. The scaled values are $d$ = 340 nm, $h$ = 595 nm, and $p$ = 435 nm [more details are provided in supplementary information (SI)].

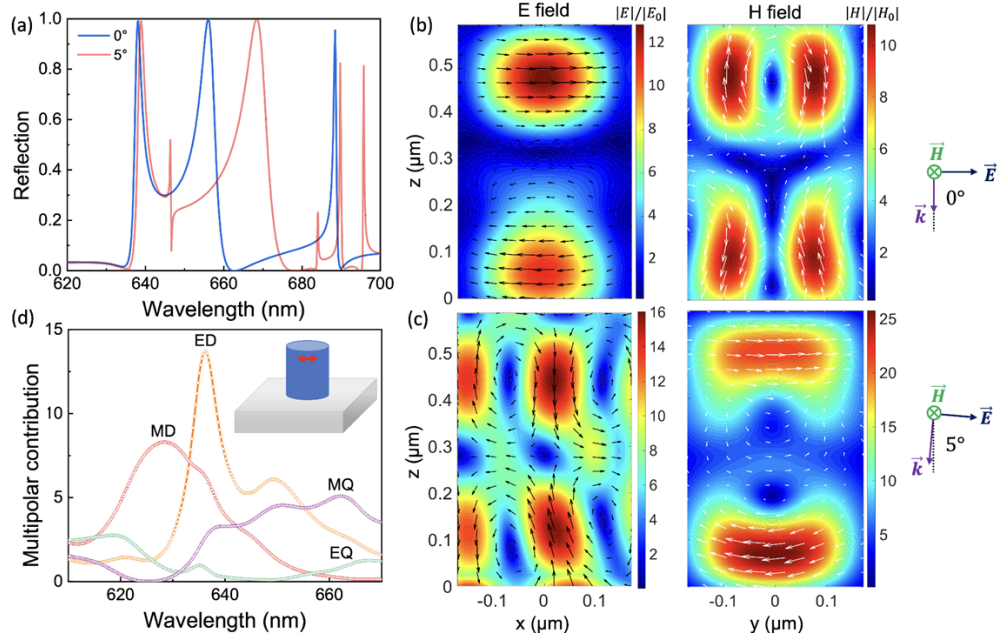

Fig. 2. **Modal analysis of the electromagnetic modes supported in the pillar array by plane wave excitation. (a)** Reflection and transmission spectra with optimized pillar array show resonances at 637 nm as well as at 655 nm and 686 nm. **(b-c)** Normalized electric (xz-plane) and magnetic field (yz-plane) distribution (E-H) inside the centermost pillar of the 2D hexagonal unit cell at 637 nm resonance for normal incident and oblique incidents respectively. **(d)** Excitation of the metastructure by placing the electric dipole at the confined electric field maxima of the pillar, i.e., 100 nm below from the top of the pillar. Different multipoles get excited, and the multipolar contribution is shown here through multipole expansion method around 637 nm. In the **inset**, the unit diamond pillar on top of glass substrate represents the centermost pillar of the hexagonal array and the point dipole is denoted by the red arrow.

So far, we assumed normally incident plane wave excitation from outside onto the nanopillars and ensured the presence of multimode. However, the design is for a dipole emitter placed *inside* the pillar and emitting light at *all* angles. To ensure multimode excitation will be present for the dipole emitter as well, as a final step in optimization, we computed the frequency- and angle- dependent reflection and transmission spectra for an incident plane wave. The spectra displayed resonant features at several wavelengths. We noticed a cross-over between supported ED and MD modes near 637 nm resonance at the incident angle of ~5° and ~15° respectively [Figure S7(a-b)]. To confirm the multimode at 637 nm, we calculated the reflection spectra [Figure 2a] and electric and magnetic field distribution for normal [Figure 2b] and 5° off normal [Figure 2c] incident light. We clearly see that ED for normal and MD at 5° off normal incidence of 637 nm plane wave. The E- and H- field profiles for higher incident angle (15°) are shown in the SI [Figure S7(c-d)] for the completeness, which demonstrates the presence of both ED and MD at the 637 resonances.

Having optimized and characterized the modes of the structure, we studied the emission characteristics of a single NV emitter, represented as a point electric dipole [27], placed in one nanopillar in the middle of an array of 37 nanopillars. Periodic boundary condition is not used and this case mimics one emitter containing pillar in a finite array of pillars. For maximum Purcell enhancement, we placed the dipole at the electric field maxima (~ 100 nm from the top surface) of the pillar and oriented along the field direction. This is similar to the [111] oriented diamond pillar with NV dipole in perpendicular orientation [28].

| | |
|---|---|
| $P_d$ | = Power emitted from the dipole emitter when placed in homogeneous diamond medium |
| $P$ | = Power emitted from the same dipole emitter when placed in diamond metaelement |
| $P_\theta$ | = Power emitted from the dipole in $\theta$ steradian around the normal in the forward direction |
| $\eta_{FE}$ | = Forward efficiency = $P_{2\pi}/P$ |
| $\eta_{DE}(\theta)$ | = Directional efficiency = $P_\theta/P_{2\pi}$ |
| $\eta(\theta)$ | = Total directional efficiency = $P_\theta/P_d$ |
| $F(\theta)$ | = Photon count enhancement factor = NV PL spectrum * $\eta_\theta$ |

Table 1: **Definition of various performance terms.**

As expected, the multipoles decomposition of the emitted dipole field at 637 nm shows emission predominantly into the ED and the MD modes [Figure 2(d)]. The interference between the radiation patterns associated with the two modes helps to achieve Kerker condition, i.e., directional emission in forward direction [21]. Due to the moderate refractive index, the higher order magnetic quadrupoles mainly MQ is also excited. The significant contribution of MQ at the resonance enhance the scattering in lateral directions and into the substrate, makes the system leakier [29,30].

This outlined approach can be similarly implemented for efficient single photon emission at different wavelengths or ZPLs associated with other colour centers present in any wide bandgap non-absorbing dielectrics like hBN, SiC and others.

## 3. Results and discussion

Various performance evaluation terms are defined in Table 1. The Purcell factor ($Q$) is the ratio between the total power emitted with metasurface to that without metasurface [31]. With our design we are able to achieve a $Q$ of 2.75 (red line, Figure 3a) and forward efficiency, $\eta_{FE}$ of 40% at the 637 nm (green line, Figure 3a). The periodic arrangement of the nanopillars confine the electromagnetic field through Mie resonance and helps to increase $Q$. The narrow forward directionality, Figure 3b, and diffused backward emission, Figure 3c, at ZPL is achieved through the Kerker condition. Most of the emission is confined to 20° angle in the forward direction. In the backward direction most of the light gets scattered to the glass substrate (seen as faint spots in the polar plot).

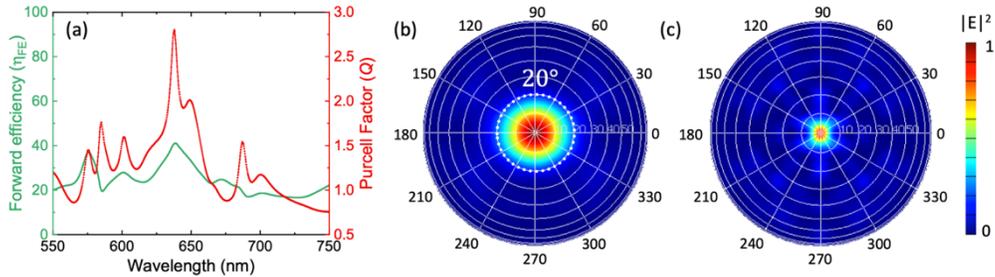

Fig. 3. **Emission from an electric dipole placed in the optimized diamond metasurface.** (a) Purcell factor (red line) and extraction efficiency (green line) are shown for the broad wavelength and the significance enhancement is observed at 637 nm for both cases. (b) Forward and (c) backward far field projection of 637 nm light of x-y planes few wavelengths far from top of the pillar and bottom of the pillar. The far-field profile is calculated by projecting the electric fields on a hemispherical surface located 1 meter away. The colour bar represents the normalized electric field intensity.

By further increasing the number of pillars to 61, $\eta_{FE}$ increases to 50% and the forward emission gets confined in much smaller angle ~ 10° [Figure S8]. The complete suppression of emission in backward direction did not happen because of the presence of quadrupole modes.

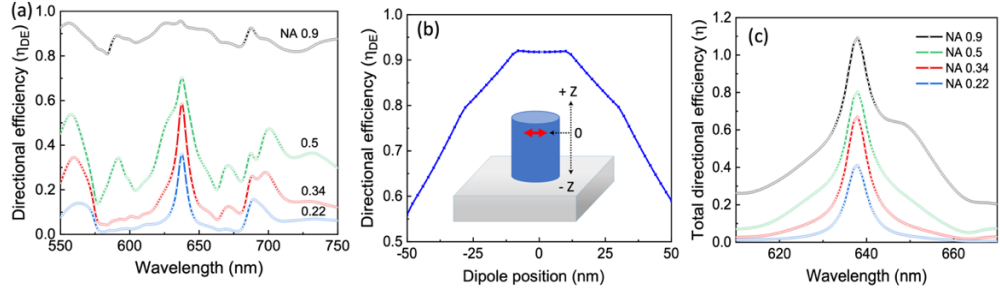

Fig. 4. **Light collection from the dipole emitter. (a)** Directional efficiency corresponds to different objective lens (denoted by NA) for our designed diamond based metasurface. **(b)** ZPL directional efficiency with respect to the position shift of the dipole. The optimized position of the dipole or the electric field maxima is represented as 0 and shift in dipole position is taken along the z axis. **(c)** Total directional efficiency for different numerical aperture around 637 nm ZPL. $NA = n * \sin(\theta)$, where n is the refractive index of air.

The directional efficiency $\eta(\theta)$ is an effective measure of quality of the design and quantified as the amount of light gets collected in a solid angle $\theta$ from the amount of available light for the collection in $2\pi$ angle. $\theta$ is related to the NA of the collection lens. We see from Figure 4a, for NA of 0.9, $\eta_{DE}$ (64°) is very large and is ~ 0.9. However, owing to broad angle of collection, radiation at all wavelengths is collected. By decreasing NA (and thus collection angle) we see that off-resonant wavelengths are rejected more, resulting in narrower peak around 637 nm. Quantitatively, the directional efficiency at ZPL is > 0.7 within the solid angle of 30°(NA of 0.5), > 0.6 within the solid angle of 20° (NA of 0.34), and ~0.4 within a solid angle of 13° (NA= 0.22) with spectral broadening of only 5 nm. As can be seen we need only a low NA collection in our case, which is especially well-suited for fiber-optic collection.

Since precise positioning of the defect is not possible in the fabrication, we evaluated the effect of emitter position on the directional efficiency. Figure 4(b) shows $\eta_{DE}$ (64°) at ZPL for various emitter positions. We see that $\eta_{DE}$ is more than 0.6 even if the dipole position is shifted $\pm$ 50 nm axially from the optimized location and is more than 0.8 when the position changes only by $\pm$ 25 nm.

The total directional efficiency $\eta(\theta)$, which includes the enhancement from $Q$ and reduction from $\eta_{FE}$, is plotted for various $\theta$ and wavelengths in Figure 4c. Owing to $Q>1$, $\eta(\theta)$ is more than 1 for larger NA.

To understand the contribution to the enhanced directional efficiency, we study four cases—bulk diamond, single diamond nanopillar on diamond substrate, single diamond nanopillar on glass substrate and array of diamond nanopillars (current work)—as shown in Figure 5a. All cases contain only a single emitter and NA of 0.9 is assumed.

In case of bulk (red line in Figure 5b), the $\eta(64°)$ is only 0.03-0.04. In other words, of 100 ZPL photons emitted at the source, only 3-4 photons are collected with NA of 0.9. With a single diamond nanopillar, the efficiency (blue, Figure 5b) increases to ~ 0.4, meaning 40 ZPL photons are collected. With glass substrate, the number (green in Figure 5b) increases to ~ 0.6 due to the total internal reflection from the diamond-glass interface. With our design, the brightness (black, Figure 5b) increases to 1.2, or equivalently 120 photons for every 100 photons emitted from the bulk diamond. Once again, $Q$ increases the brightness (number of emitted photons).

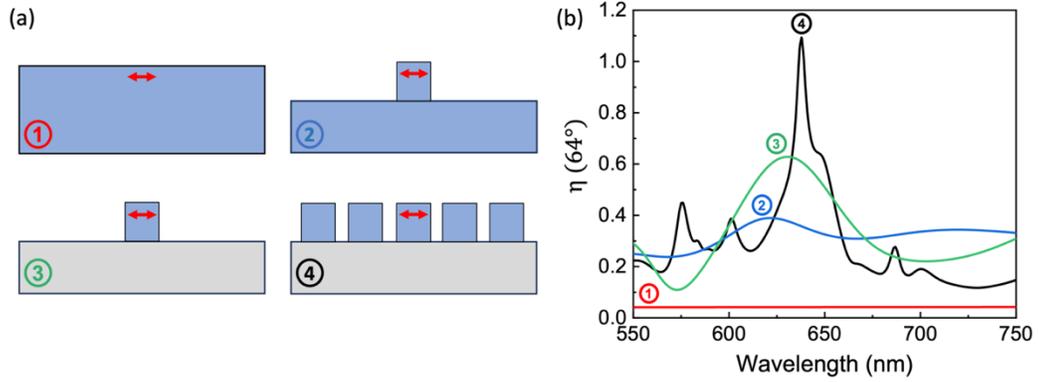

Fig. 5. **Comparison between different scenario with our diamond metasurface.** **(a)** 2D drawing of the different cases: 1. Bulk diamond substrate, 2. single diamond nanopillar on top of diamond substrate, 3. single diamond pillar on top of glass substrate, 4. our proposed designed with arrays of diamond nanopillar on top of glass substrate. The colour codes of diamond and glass are blue and light grey respectively. The red arrow denotes the electric dipole of wavelength 637 nm, perpendicular to the pillar axis and situated 100 nm below from the top surface of each case. **(b)** Total directional efficiency is shown for different cases with NA of 0.9 (or θ =64°) over the long wavelength ranges.

Figure 6(a) shows the predicted photon count enhancement $F_θ$ (for θ =13°, or NA 0.22) over the emission at ZPL from the bulk diamond. We used the measured PL spectrum [Figure S1] from the NV center in the bulk, normalized it to the value at ZPL and multiplied it by the calculated η(13°) to get $F_θ$. In essence, photon number at ZPL from the metasurface is $F_θ$ times that collected from the bare diamond surface. We see that enhancement is nearly 160. Inset shows the 2D radiation pattern for both bulk and metasurface along the xz- and yz- plane, for the electric dipole placed along the x axis. We predict a maximum radiation enhancement of ~ 200x and mostly in the forward direction with the designed metasurface.

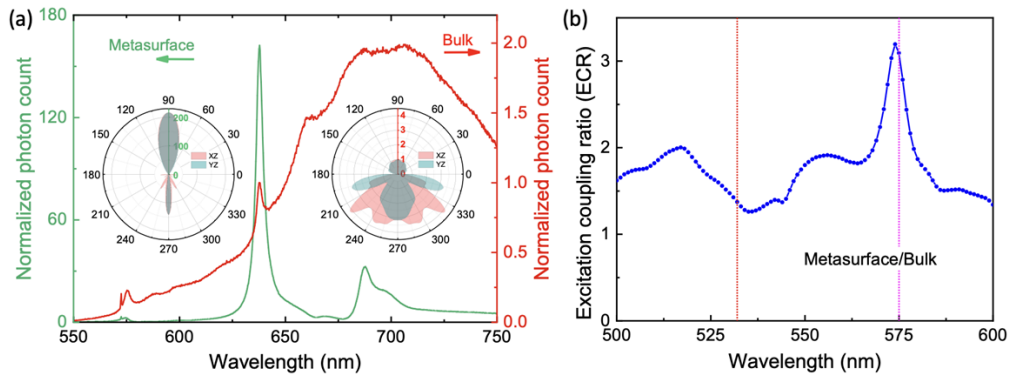

Fig. 6. **ZPL brightness in diamond metasurface compared to bulk.** **(a)** Normalized phonon counts from our designed diamond metasurface w.r.t the bulk ZPL count. (Inset) Normalized 2D radiation pattern of metasurface (left) and bulk diamond (right) at ZPL in the xz- and yz- plane. **(b)** Excitation coupling ratio between diamond metasurface and bulk with broadband Gaussian source.

Further, the metasurface behaves like an antireflection or photon trapping layer for shorter wavelengths pump, thus increasing the coupling to NV centers which results in an increased effective efficiency. The excitation coupling ratio (ECR), shown in Figure 6(b), measures how wavelength-dependent pump couples to the NV centers in the metasurface compared to that in the bulk diamond. The ratio is calculated from the absorbed light within a small volume of $(340 \times 340 \times 100)$ nm$^3$ around the emitter for both cases, where 340 nm is the diameter of the pillar and 100 nm is the height of the electric field confined region inside the pillar. The absorption is calculated using the measured absorption coefficient value [32]. A broadband gaussian light source for excitation with NA 0.22 is assumed. We see from Figure 6b that an ECR of 1.37 at the most common pump wavelength of 532 nm. By multiplying photon enhancement factor ($F_\theta$) by ECR we get an effective ZPL photon enhancement of ~222. Instead, if a 575 nm pump is used, the enhancement factor can reach ~500.

We make a few observations before concluding. *First*, the exploitation of Mie resonance and Kerker condition for emission enhancement has been studied by several others where the single photon source is either a NV center embedded in Si metasurface [33], Si-vacancy center ($V_{Si}$) in SiC metasurface [34], or GaAs quantum dot embedded in a high index metasurface [35]. High emission was predicted in these calculations as well. Our design approach provides details needed to additionally obtain highly directional emission. Previously predicted strong emission in transverse direction along with the forward and backward emission effectively reduces the light collection efficiency. *Second*, our design considers NV center in its natural (diamond) environment, not in Si-embedded in diamond. *Third*, we consider single NV center, instead of array of NV centers as in previous studies, which is more appropriate for the development of single photon source. *Fourth*, our design requires fabrication of small area of nanopillars just around the NV center. Further, the array is on the low index (SiO$_2$ or quartz) substrate and high-quality polycrystalline diamond has been grown on these substrates [36–38]. If single crystal diamond pillars with emitter dipoles oriented in [111] direction is required, one can fabricate nanopillars of crystalline diamond grown on diamond substrate and with back-filled polymer, the array can be peeled-and-transferred to any low index substrate [39]. Hence the developed design is most realistic for fabrication and demonstration.

## 4. Conclusion

We described an approach to design a metasurface by exploiting the Mie resonance and Kerker condition for highly directional, enhanced and efficient light emission. The Mie resonance helps to achieve a higher Purcell factor and Kerker condition leads to high directionality. In this work, we designed diamond metasurface, tailored for nitrogen-vacancy (NV) defect centers in diamond, and predict ~200x to 500x improvement in the number of ZPL photons collected. Our design achieves an emission which is highly directional— predominantly within 20º cone— in the forward direction. This makes light collection more efficient, including for fiber-based collection. The predicted results are fairly stable against the position of the emitter placed in the metaelement, thus alleviating the challenging fabrication requirement of precise positioning of the defect center. Equally importantly, our design approach can be applied to enhance single photon emission also from other defects, materials. For example, by tuning the diameter and height of the pillars and the periodicity of the arrangement, we can shift the resonance to 737 nm corresponding SiV defect centers in diamond. This design algorithm can be used for other wide band dielectrics such as hBN and SiC.

## 5. Acknowledgements

This work was supported by DST, India, under QuEST program through contract number DST/ICPS/QuST/Theme-2/2019/General and by MHRD STARS research grant through contract number STARS/APR2019/396. Rituraj acknowledges support from Science & Engineering Research Board through project number SERB/EE/2022423.## 6. Conflict of interest

The authors have no conflict of interest.

contract number STARS/APR2019/396. Rituraj acknowledges support from Science & Engineering Research Board through project number SERB/EE/2022423.

## 6. Conflict of interest

The authors have no conflict of interest.

## 7. References:

1. I. Aharonovich, D. Englund, and M. Toth, "Solid-state single-photon emitters," Nat Photonics **10**, 631–641 (2016).
2. S. Castelletto and A. Boretti, "Perspective on Solid-State Single-Photon Sources in the Infrared for Quantum Technology," Adv Quantum Technol (2023).
3. A. B. D. al jalali wal ikram Shaik and P. Palla, "Optical quantum technologies with hexagonal boron nitride single photon sources," Sci Rep **11**, (2021).
4. B. Rodiek, V. Sandoghdar, X.-L. Chu, S. Kuck, M. Lopez, S. Lindner, H. Hofer, M. Smid, C. Becher, G. Porrovecchio, and S. Gotzinger, "Experimental realization of an absolute single-photon source based on a single nitrogen vacancy center in a nanodiamond," Optica, Vol. 4, Issue 1, pp. 71-76 **4**, 71–76 (2017).
5. U. Sinha, "Single-Photon Sources," Opt Photonics News **30**, 32–39 (2019).
6. M. V. Gurudev Dutt, L. Childress, L. Jiang, E. Togan, J. Maze, F. Jelezko, A. S. Zibrov, P. R. Hemmer, and M. D. Lukin, "Quantum register based on individual electronic and nuclear spin qubits in diamond," Science (1979) **316**, 1312–1316 (2007).
7. G. Balasubramanian, I. Y. Chan, R. Kolesov, M. Al-Hmoud, J. Tisler, C. Shin, C. Kim, A. Wojcik, P. R. Hemmer, A. Krueger, T. Hanke, A. Leitenstorfer, R. Bratschitsch, F. Jelezko, and J. Wrachtrup, "Nanoscale imaging magnetometry with diamond spins under ambient conditions," Nature **455**, 648–651 (2008).
8. S. Johnson, P. R. Dolan, T. Grange, A. A. P. Trichet, G. Hornecker, Y. C. Chen, L. Weng, G. M. Hughes, A. A. R. Watt, A. Auffèves, and J. M. Smith, "Tunable cavity coupling of the zero phonon line of a nitrogen-vacancy defect in diamond," New J Phys **17**, (2015).
9. M. V. Gurudev Dutt, L. Childress, L. Jiang, E. Togan, J. Maze, F. Jelezko, A. S. Zibrov, P. R. Hemmer, and M. D. Lukin, "Quantum register based on individual electronic and nuclear spin qubits in diamond," Science (1979) **316**, 1312–1316 (2007).
10. G. Balasubramanian, I. Y. Chan, R. Kolesov, M. Al-Hmoud, J. Tisler, C. Shin, C. Kim, A. Wojcik, P. R. Hemmer, A. Krueger, T. Hanke, A. Leitenstorfer, R. Bratschitsch, F. Jelezko, and J. Wrachtrup, "Nanoscale imaging magnetometry with diamond spins under ambient conditions," Nature **455**, 648–651 (2008).
11. P. K. Shandilya, S. Flagan, N. C. Carvalho, E. Zohari, V. K. Kavatamane, J. E. Losby, and P. E. Barclay, "Diamond Integrated Quantum Nanophotonics: Spins, Photons and Phonons," Journal of Lightwave Technology **40**, 7538–7571 (2022).
12. D. Riedel, I. Söllner, B. J. Shields, S. Starosielec, P. Appel, E. Neu, P. Maletinsky, and R. J. Warburton, "Deterministic enhancement of coherent photon generation from a nitrogen-vacancy center in ultrapure diamond," Phys Rev X **7**, (2017).
13. T. M. Babinec, B. J. M. Hausmann, M. Khan, Y. Zhang, J. R. Maze, P. R. Hemmer, and M. Lončar, "A diamond nanowire single-photon source," Nat Nanotechnol **5**, 195–199 (2010).
14. A. Faraon, P. E. Barclay, C. Santori, K. M. C. Fu, and R. G. Beausoleil, "Resonant enhancement of the zero-phonon emission from a colour centre in a diamond cavity," Nat Photonics **5**, 301–305 (2011).
15. A. Faraon, C. Santori, Z. Huang, V. M. Acosta, and R. G. Beausoleil, "Coupling of nitrogen-vacancy centers to photonic crystal cavities in monocrystalline diamond," Phys Rev Lett **109**, 033604 (2012).
16. S. Johnson, P. R. Dolan, and J. M. Smith, "Diamond photonics for distributed quantum networks," Prog Quantum Electron **55**, 129–165 (2017).
17. T. Schröder, S. L. Mouradian, J. Zheng, M. E. Trusheim, M. Walsh, E. H. Chen, L. Li, I. Bayn, and D. Englund, "Quantum nanophotonics in diamond [Invited]," Journal of the Optical Society of America B **33**, B65 (2016).
18. Y. Kivshar and A. Miroshnichenko, "Meta-Optics with Mie Resonances," *Optics and Photonics News* 28.1 (2017): 24-31.